\documentstyle[psfig,times]{jaa}
\begin{document}

\title[Viscous and Resistive ADAF]{Self-Similar Solutions for Viscous and Resistive ADAF}
\author[K. Faghei]%
{Kazem Faghei \thanks{e-mail: kfaghei@du.ac.ir}
\\
School of Physics, Damghan University, Damghan, Iran}
\pubyear{2011}
\date{Received 2011 January; accepted 2011 April}
\maketitle
\label{firstpage}
\begin{abstract}
 In this paper, the self-similar solution of resistive advection dominated accretion flows (ADAF) in the presence of a pure azimuthal magnetic field is investigated. The mechanism of energy dissipation is assumed to be the viscosity and the magnetic diffusivity due to turbulence in the accretion flow. It is assumed that the magnetic diffusivity and the kinematic viscosity are not constant and vary by position and $\alpha$-prescription is used for them. In order to solve the integrated equations that govern the behavior of the accretion flow, a self-similar method is used. The solutions show that the structure of accretion flow depends on the magnetic field and the magnetic diffusivity. As, the radial infall velocity and the temperature of the flow increase, and the rotational velocity decreases. Also, the rotational velocity for all selected values of magnetic diffusivity and magnetic field is sub-Keplerian. The solutions show that there is a certain amount of magnetic field that the rotational velocity of the flow becomes zero. This amount of the magnetic field depends on the gas properties of the disc, such as adiabatic index and viscosity, magnetic diffusivity, and advection parameters. The solutions show the mass accretion rate increases by adding the magnetic diffusivity and in high magnetic pressure case, the ratio of the mass accretion rate to the Bondi accretion rate decreases as magnetic field increases. Also, the study of Lundquist and magnetic Reynolds numbers based on resistivity indicates that the linear growth of magnetorotational instability (MRI) of the flow decreases by resistivity. This property is qualitatively consistent with resistive magnetohydrodynamics (MHD) simulations.
\end{abstract}


\newpage
\section{Introduction}

The standard geometrically thin, optically thick accretion disc model can successfully explain most of the observational features in active galactic nuclei (AGN) and X-ray binaries (Shakura \& Sunyaev 1973).  In the standard thin model, the motion of the matter in the accretion disc is nearly Keplerian, and the gravitational energy released in the disc is radiated away locally. During recent years, another type of accretion flow has been studied in which it is assumed that energy released through dissipation processes in the disc may be trapped within accreting gas, and only the small fraction of the energy released in the accretion flow is radiated away due to inefficient cooling, and most of the energy is stored in the accretion flow and advected to the central object. This kind of accretion flow is known as advection-dominated accretion flow (ADAF). The basis ideas of ADAF models have been developed by a number of researchers (e.g. Ichimaru 1977; Rees et al. 1982; Narayan \& Yi 1994, 1995; Abramowicz et al. 1995; Ogilvie 1999; Blandford \& Begelman 1999). 

It seems that accretion discs, whether in star-forming regions, in X-ray binaries, in cataclysmic variables, or in the centers of active galactic nuclei, are likely to be threaded by magnetic fields. Consequently, the role of the magnetic fields on ADAFs has been analyzed in detail by a number of investigators (Bisnovatyi-kogan \& Lovelace 2001; Akizuki \& Fukue 2006, hereafter AF06; Shadmehri 2004, hereafter Sh04; Ghanbari et al., 2007; Bu, Yung, \& Xie 2009; Khesali \& Faghei 2009, hereafter KF09). The existence of the toroidal magnetic field has been proven in the outer regions of the discs of young stellar objects (YSOs; Aitken et al. 1993; Wright et al. 1993) and in the Galactic center (Novak et al. 2003; Chuss et al. 2003). Thus, considering the accretion discs with a toroidal
magnetic field have been studied by several authors (AF06; Begelman \& Pringle 2007; Khesali \& Faghei 2008, hereafter KF08; Bu, Yung, \& Xie 2009; KF09).

The resistive diffusion of magnetic field is important in some systems, such as the protostellar discs (Stone et al. 2000; Fleming \& Stone 2003), discs in dwarf nova systems (Gammie \& Menou 1998), the discs around black holes (Kudoh \& Kaburaki 1996), and Galactic center (Melia \& Kowalenkov 2001; Kaburaki et al. 2010). Also, two and three dimensional simulations of local shearing box have shown that resistive dissipation is one of the crucial processes that determines the saturation amplitude of the magnetorotational instability (MRI). As, linear growth rate of MRI can be reduced significantly because of the suppression by ohmic dissipation (Fleming et al. 2000; Masada \& Sano 2008).

AF06 proposed a self-similar advection-dominated accretion flow that the disc plasma is highly ionized, so they assumed that resistivity of the plasma is zero, and only viscosity is due to turbulence and dissipation in the disc. However, recent works represent importance of magnetic diffusivity in accretion discs (e.g. Kuwabara et al. 2000; Kaburaki 2000 ;Kaburaki et al. 2002; Sh04; Ghanbari et al. 2007; Krasnopolsky et al. 2010; Kaburaki et al. 2010). Sh04 studied a quasi-spherical accretion flow that dominant mechanism of energy dissipation was assumed to be the magnetic diffusivity due to turbulence in the accretion flow and the viscosity of the fluid was completely neglected. The main focus of Sh04 was nonrotating accretion flow and ignored from toroidal magnetic field. Also, Sh04 studied induction equation of magnetic field in a steady state that is not according to anti-dynamo theorem (e.g. Cowling 1981) and is usefull only in particular systems where the magnetic dissipation time is very long, much longer than the age of the system. Ghanbari, Salehi, \& Abbassi (2007) considered an axisymmetric, rotating, isothermal steady accretion flow, which contains a poloidal magnetic field of the central star and from toroidal magnetic field of the flow neglected. They assumed that the mechanisms of energy dissipation are the turbulence viscosity and magnetic diffusivity due to the magnetic field of the central star. They explored the effect of viscosity on a rotating disc in the presence of constant magnetic diffusivity. Similar to Sh04 they considered the flow in balance between escape and creation of the magnetic field, and ignored from toroidal component of magnetic field. 

As mentioned the observational evidences and the MHD simulations express that the toroidal component of magnetic field and the magnetic diffusivity are important in accretion discs. Thus in this paper by using AF06 technique we will study the influence of presence of toroidal component of magnetic field in a viscous and resistive accreting gas, and investigate the role of non-constant magnetic diffusivity in systems that escaping and creating of magnetic field are not balanced. We will show that the present model from some aspects is qualitatively consistent with the observational evidences and the resistive MHD simulation results. The paper is organized as follows. In section 2, the basic equations of constructing a model for quasi-spherical magnetized advection dominated accretion flow will be defined. In section 3, self-similar method for solving equations which govern the behavior of the accreting gas was utilized. The summary of the model will appear in section 4.

\section{Basic Equations}
We use spherical coordinate $(r,\theta,\varphi)$ centered on the accreting object and make the following standard assumptions:
\begin{itemize}
  \item [(i)] The flow is assumed to be steady and axisymmetric $ \partial_{t}=\partial_{\varphi}=0$, so all flow variables are a function of only $r$ ;
  \item [(ii)] The magnetic field has only an azimuthal component; 
  \item [(iii)] The gravitational force on a fluid element is characterized by the Newtonian potential of a point mass, $\Psi=-{GM_{*}}/{r}$,   with $G$ representing the gravitational constant and $M_{*}$ standing for the mass of the central star;
  \item [(iv)] The equations written in spherical coordinates are considered in the equatorial   plane $\theta=\pi/2$ and  terms with any $\theta$ and $\varphi$ dependence are neglected (Ogilvie 1999; KF09).
  \item [(v)] For the sake of simplicity, the self-gravity and general relativistic effects have been neglected;
  
\end{itemize}
The behavior of such system can be analyzed by magnetohydrodynamics (MHD) equations. The general MHD equations are written as follows:

\begin{equation}
\frac{\partial \rho}{\partial t}+\nabla\cdot(\rho{\bf v})=0,
\end{equation}
\begin{equation}
\rho
\left[
\frac{\partial{\bf v}}{\partial t}+({\bf v}\cdot\nabla){\bf v}
\right]
=-\nabla p_{gas} -\rho \nabla\Psi + \frac{1}{4\pi}{\bf
J}\times\bf{B}+{\bf{F}}_{vis},
\end{equation}
\begin{equation}
{\bf J}=\nabla\times {\bf B}
\end{equation}
\begin{equation}
\frac{\partial{\bf B}}{\partial t}=\nabla\times\left({\bf v}\times{\bf
B }-\eta {\bf J}\right),
\end{equation}
\begin{equation}
\rho\left[\frac{1}{\gamma-1}\frac{d}{d t}\left( \frac{p_{gas}}{\rho}\right)+\left( \frac{p_{gas}}{\rho}\right) (\nabla\cdot{\bf
v})\right]=Q_{\rm diss}-Q_{\rm cool}\equiv fQ_{\rm diss}, 
\end{equation}
\begin{equation}
\nabla \cdot {\bf B}=0,
\end{equation}
where $\rho$, $\mathbf{v}$, and $p_{gas}$ are the density, the velocity field, and the gas pressure, respectively; $\mathbf{F}_{vis}$ is the viscous force per unit volume; $\mathbf{B}$ is the magnetic field; ${\bf J}$ is the current density and $\eta$ represents the magnetic diffusivity; $\gamma$ is the adiabatic index; The term on the right hand side of the energy equation, $Q_{\rm diss}$, is the rate of heating of the gas by the dissipation and $Q_{\rm cool }$ represents the energy loss through radiative cooling. The advection factor, $f$ ($0 \leq f \leq 1$), describes the fraction of the dissipation energy which is stored in the accretion flow and advected into the central object rather than radiated away. The advection factor of $f$ in general depends on the details  of the heating and cooling mechanism and will vary with postion (e.g. Watari 2006, 2007; Sinha et al. 2009). However, we assume a constant $f$ for simplicity. Clearly, the case $f=1$ corresponds to the extreme limit of no radiative cooling and in the limit of efficient cooling, we have $f=0$.

Under the assumptions (i)-(v), the approximation of quasi-spherical symmetry, the equations (1)-(6) become

\begin{equation}
\frac{1}{r^{2}}\frac{d}{d r}(r^{2}\rho v_{r})=\dot{\rho},
\end{equation}

\begin{equation}
v_{r}\frac{d
v_{r}}{d r}
  +\frac{1}{\rho}\frac{d p}{d r}
  +\frac{GM_{*}}{r^{2}}
   =r\Omega^{2}-\frac{B_{\varphi}}
     {4 \pi r\rho}\frac{d }{d r}(rB_{\varphi}),
\end{equation}

\begin{equation}
\rho v_{r}\frac{d}{d r}(r^{2}\Omega)=
  \frac{1}{r^{2}}\frac{d}{d r}\left[\nu\rho r^{4}\frac{d \Omega}{\partial r}\right],
\end{equation}

\begin{equation}
 \nonumber \frac{1}{\gamma-1}\left[v_r\frac{d p}{d r}+
\frac{\gamma p}{r^2}\frac{d}{d
r}\left( r^2 v_r\right)\right]= f Q_{diss},
\end{equation}

\begin{equation}
\frac{1}{r}\frac{d
}{d r}\left[r v_{r} B_{\varphi}-\eta\frac{d
}{d r}(r B_{\varphi})\right]=\dot{B}_{\varphi}.
\end{equation}
Here $v_r$ the radial velocity, $\dot\rho$ the mass-loss rate per unit volume, $\Omega$ the angular velocity, $B_{\varphi}$ the toroidal component of magnetic field, $\dot{B}_{\varphi}$ is the field escaping/creating rate due to a magnetic instability or dynamo effect, $\nu$ the kinematic viscosity coefficient.  We assume both of the kinematic coefficient of viscosity and the magnetic diffusivity due to turbulence in the accretion flow, it is reasonable to use these parameters in analogy to the $\alpha$-prescription of Shakura \& Sunyaev (1973) for the turbulent,

\begin{equation}
   \nu = \alpha \frac{p_{gas}}{\rho\Omega_{K}}(1+\beta)^{1-\mu},
\end{equation}
\begin{equation}
   \eta = \eta_0 \frac{p_{gas}}{\rho\Omega_{K}}(1+\beta)^{1-\mu}
\end{equation}
where $\Omega_K=({GM_*}/{r^3})^{1/2}$ is the Keplerian angular velocity. Narayan \& Yi (1995) used a similar form for kinematic coefficient of viscosity, i.e. $\nu=\alpha (p_{gas}/\rho\Omega_{K})$, also Sh04 applied a similar form for magnetic diffusivity, i.e. $\eta = \eta_0 (p_{gas}/\rho\Omega_{K})$. Thus in comparison to Narayan \& Yi (1995) and Sh04 prescriptions, we are using the above equations for the kinematic coefficient of viscosity $\nu$ and the magnetic diffusivity $\eta$. The parameters of $\alpha$ and $\eta_0$ are assumed to be positive constants, and we assume that $\alpha$, $\eta_0\le 1$ (Campbell 1999; Kuwabara et al. 2000; Sh04; King et al. 2007).  The parameter of $\beta[=p_{mag}/p_{gas}]$ is the degree of magnetic pressure, $p_{mag}=B^{2}_{\varphi}/8\pi$, to the gas pressure. Since, we will apply steady self-similar method for solving system equations, this parameter will be constant throughout the disc, but really this parameter varies by position (see, KF08, KF09). The parameter of $\mu$ is a constant and states importance degree of total pressure in the kinematic viscosity and the magnetic diffusivity. Clearly, the case of $\mu=1$ corresponds to traditional $\alpha$-model and in case of $\mu=0$ total pressure is used. Note the kinematic coefficient of viscosity and the magnetic  diffusivity are not costant and depend on the physical quantities of the flow. we will show the quantities of $\nu$ and $\eta$ in our self-similar solution vary with radius $r^{1/2}$.

The ratio of the kinematic coefficient of viscosity to the magnetic diffusivity is defined by the magnetic Prandtl number, $P_m=\nu/\eta$. By using equations (12) and (13) the magnetic Prandtl number  in present model is $P_m=\alpha/\eta_0$. We will consider conditions that $P_m=\infty$ (in case that  magnetic diffusivity is zero), $P_m \ge 1$, and $P_m < 1$. 

For the heating term in equation (10), $Q_{diss}$, we use two sources of dissipation: the viscous and  resistive dissipations. Thus, for $Q_{diss}$ we can write
   
\begin{equation}
   Q_{diss}=\nu\rho r^2 \left(\frac{d \Omega}{d r}\right)^2+\frac{\eta}{4\pi}J^2
\end{equation}
where first term is related to viscous dissipation and second term is related to resistive dissipation. By converting the gas pressure and the magnetic pressure in terms of sound speed ($c^{2}_s=p_{gas}/\rho$) and Alfv\'{e}n speed ($c^{2}_A=B^{2}_{\varphi}/4\pi\rho$), and by using equations (12)-(14) for the equations (7)-(10), we can write

\begin{equation}
\frac{1}{r^{2}}\frac{d}{d r}(r^{2}\rho v_{r})=\dot{\rho},
\end{equation}

\begin{equation}
v_{r}\frac{d
v_{r}}{d r}
  +\frac{1}{\rho}\frac{d }{d r}(\rho c^2_s)
  +\frac{GM_{*}}{r^{2}}
   =r\Omega^{2}-\frac{c^2_A}{r}-\frac{1}
     {2 \rho}\frac{d }{d r}(\rho c^2_A),
\end{equation}

\begin{equation}
\rho v_{r}\frac{d}{d r}(r^{2}\Omega)=
  \frac{\alpha}{r^{2}}\frac{d}{d r}\left[\frac{\rho c^2_s}{\Omega_K}(1+\beta)^{1-\mu}r^4\left(\frac{d \Omega}{d r}\right)\right],
\end{equation}

\begin{equation}
 \nonumber \frac{1}{\gamma-1}\left[v_r\frac{d}{d r}(\rho c^2_s)+
\frac{\gamma \rho c^2_s}{r^2}\frac{d}{d r}\left( r^2 v_r\right)\right]= f Q_{diss},
\end{equation}

\begin{equation}
\frac{1}{r}\frac{d
}{d r}\left[\sqrt{4\pi\rho c^2_A}\left(r v_{r} -\frac{\eta_0 (1+\beta)^{1-\mu}}{4\rho\beta\Omega_K}\frac{d
}{d r}(r^2 \rho c^2_A )\right)\right]=\dot{B}_{\varphi},
\end{equation}
where
\begin{equation}
   Q_{diss}=\frac{(1+\beta)^{1-\mu}}{\Omega_{K}}
\left[
\alpha r^2 \rho c^2_{s} \left(\frac{d \Omega}{d r}\right)^2+\frac{\eta_0}{8 r^4 \rho \beta } \left(\frac{d
}{d r}(r^2 \rho c^2_A )\right)^2
\right].
\end{equation}

\section{Self-Similar Solutions}
We seek self-similar solutions in the following forms:

\begin{equation}
v_r=-c_1 \alpha \sqrt{\frac{G M_*}{r}}
\end{equation}

\begin{equation}
\Omega=c_2\sqrt{\frac{G M_*}{r^3}}
\end{equation}

\begin{equation}
c^2_s=c_3\frac{G M_*}{r}
\end{equation}

\begin{equation}
c^2_A=\frac{B^2_{\varphi}}{4\pi\rho}=2 \beta c_3\frac{G M_*}{r}
\end{equation}
where coefficients $c_1$, $c_2$, and $c_3$ are determined later. we assume a power-law relation for density

\begin{equation}
\rho(r)=\rho_0 r^s
\end{equation}
where $\rho_0$ and $s$ are constant. By using above self-similar quantities, the mass-loss rate and the field escaping/creating rate must have the following form:

\begin{equation}
\dot{\rho}=\dot{\rho}_0 r^{s-3/2}
\end{equation} 

\begin{equation}
\dot{B}_{\varphi}=\dot{B}_0 r^{\frac{s-4}{2}}
\end{equation}
where $\dot{\rho}_0$ and $\dot{B}_0$ are constant. 

Substituting the above solutions in the continuity, momentum, angular momentum, energy, and induction equations [(15)-(20)], we can obtain the following relations:

\begin{equation}
\dot{\rho}_0=-\left(s+\frac{3}{2}\right) \alpha \rho_0 c_1 \sqrt{G M_*},   
\end{equation}

\begin{equation}
-\frac{1}{2}c^2_1 \alpha^2 + 1 - c^2_2 + c_3 \left(s-1+\beta (1+s)\right)=0    ,
\end{equation}

\begin{equation}
c_1=3 (s+2)(1+\beta)^{1-\mu} c_3, 
\end{equation}

\begin{equation}
-\frac{1}{\gamma-1}\alpha c_1 (2s-2+3 \gamma)=\frac{1}{2} f (1+\beta)^{1-\mu} \left(
9 \alpha c^2_2 +2 \eta_0 \beta c_3 (1+s)^2
\right), 
\end{equation} 

\begin{equation}
\dot{B}_0=-\frac{s}{2} G M_* \sqrt{2\pi\rho_0\beta c_3}\left(
2 c_1 \alpha + \eta_0 c_3 (1+s) (1+\beta)^{1-\mu}
\right).  
\end{equation} 

Above equations express for $s=-3/2$, there is no mass loss, while for $s > -3/2$ mass loss (wind) exists. The escape and creation of magnetic fields are balanced in $s=-2+(3\eta_0)/(6\alpha+\eta_0)$ that $\dot{B}_0$ becomes zero, solving it in $s=-3/2$ (no wind) implies that $\eta_0=(6/5)\alpha$. This amount of $\eta_0$ corresponds to the magnetic Prandtl number $5/6$. Thus, when the flow has $s=-3/2$ and $\eta_0=(6/5)\alpha$, we expect the escape and creation of magnetic field are balanced and there is no mass loss. In $\eta_0=0$ for balance of escape and creation of magnetic field, we have $s=-2$. In this paper, only case of no wind ($s=-3/2$) is considered that $\dot\rho_0=0$ and $\dot{B}_{\varphi}\propto r^{-11/4}$ , we note $\dot{B}_{\varphi}$ in $P_m=5/6$ will be zero, too. The equations (29)-(31) imply that for given $\alpha$, $\eta_0$, $\beta$, $s$, and $f$ form a closed set of equations of $c_1$, $c_2$, and $c_3$ which will determine behavior of the accretion flow.  

\subsection{Analysis}
By using the equations (29)-(31), the coefficients of $c_i$ have the following forms  

\begin{equation}
 c_1=\frac{1}{2\alpha^2}
\left(
D_4+\sqrt{D^2_4+8 \alpha^2}
\right),
\end{equation} 

\begin{equation}
 c^2_2=-\frac{2}{9} c_1 D_1 D_2,
\end{equation} 

\begin{equation}
c_3=\frac{1}{3(s+2)} c_1 D_1,
\end{equation} 
\
where
\
\begin{equation}
D_1=\frac{1}{(1+\beta)^{1-\mu}},
\end{equation} 

\begin{equation}
D_2=\frac{(2s-2+3\gamma)}{(\gamma-1)f}+\frac{\eta_0 \beta}{3\alpha} \frac{(1+s)^2}{(s+2)},
\end{equation} 

\begin{equation}
D_3=\frac{s-1+\beta(1+s)}{s+2},
\end{equation} 

\begin{equation}
D_4=\frac{2}{3} D_1 
\left(
\frac{2}{3}D_2+D_3
\right).
\end{equation} 
The obtained results imply that the model is parametrized by the ratio of specific heat, $\gamma$, the standard viscous parameter, $\alpha$, the magnetic diffusivity parameter, $\eta_0$, the degree  of magnetic pressure to gas pressure, $\beta$, the advection parameter, $f$, and the mass-loss rate parameter, $s$.

The equation (37) implies that the value of $D_2$ for a value of $\beta$  will be zero. From equations (22) and (34) we can write $\Omega \propto c_2^2 \propto D_2$. Thus we conclude for a value of $\beta$ that we define it $\beta_b$ (braking $\beta$), the angular velocity will be zero. Over of $\beta_b$, $c_2^2$ becomes negative that it is not physical. By solving $D_2=0$ for the $\beta$ parameter, we can write

\begin{equation}
 \beta_b=-\frac{3\alpha}{f\eta_0 }\frac{(s+2)}{(s+1)^2}\frac{(2s-2+3\gamma)}{(\gamma-1)} .
\end{equation}   
In the case of no mass-loss, $s=-3/2$, we can write
\begin{equation}
 \beta_b=\frac{18\alpha}{f\eta_0 }\left(\frac{5/3-\gamma}{\gamma-1}\right).
\end{equation}
The above equation in terms of the magnetic Prandtl number becomes
\begin{equation}
 \beta_b=18 (\frac{P_m}{f})\left(\frac{5/3-\gamma}{\gamma-1}\right).
\end{equation}
The equations (40)-(42) express that the amount of the $\beta_b$ parameter depends on $f$, $P_m$, and $\gamma$. As $\beta_b$ decreases by adding the advection degree, while increases by adding the magnetic Prandtl number. In a flow with high conductivity, that $P_m$ is very large, $\beta_b$ becomes very large. Also in the flow with high resistivity and low viscosity, $\beta_b$ will be small. Since the magnetic pressure fraction always is positive or equal to zero ($\beta \geq 0$), so in presence model $\gamma \leq 5/3$.

Examples of the coefficients $c_i$ in two cases of $\mu=0$ and $1.0$ are shown  in figures (1) and (2) as a function of the degree of magnetic pressure to the gas pressure, ($\beta$), for different value of the magnetic diffusivity, $\eta_0$. In figures (1) and (2), $c_4$ is the viscous torque that is obtained from right hand side of equation (30), $c_5$ is total energy dissipation by the viscosity and the resistivity, and is calculated by right hand side of equation (31), and $c_6$ is the ratio of the resistive dissipation to the viscous dissipation.

\subsubsection{First Case: $\mu=0$ }
The parameter of $\mu$ has appeared in the equations of the kinematic coefficient of viscosity and the magnetic diffusivity to indicate important degree of total pressure in them. In the case of $\mu=0$, the kinematic coefficient of viscosity and the magnetic diffusivity become
\begin{eqnarray}
 \nonumber \nu=\alpha\frac{c^2_s}{\Omega_K}(1+\beta)
\\
=\alpha c_3 (1+\beta) \sqrt{G M_*}~r^{1/2}
\end{eqnarray}
\begin{eqnarray}
 \nonumber \eta=\eta_0 \frac{c^2_s}{\Omega_K}(1+\beta)
\\
=\eta_0 c_3 (1+\beta) \sqrt{G M_*}~r^{1/2}.
\end{eqnarray}

\input{epsf}
\begin{figure}[!ht]
\begin{center}
\centerline{{\epsfxsize=7.5cm\epsffile{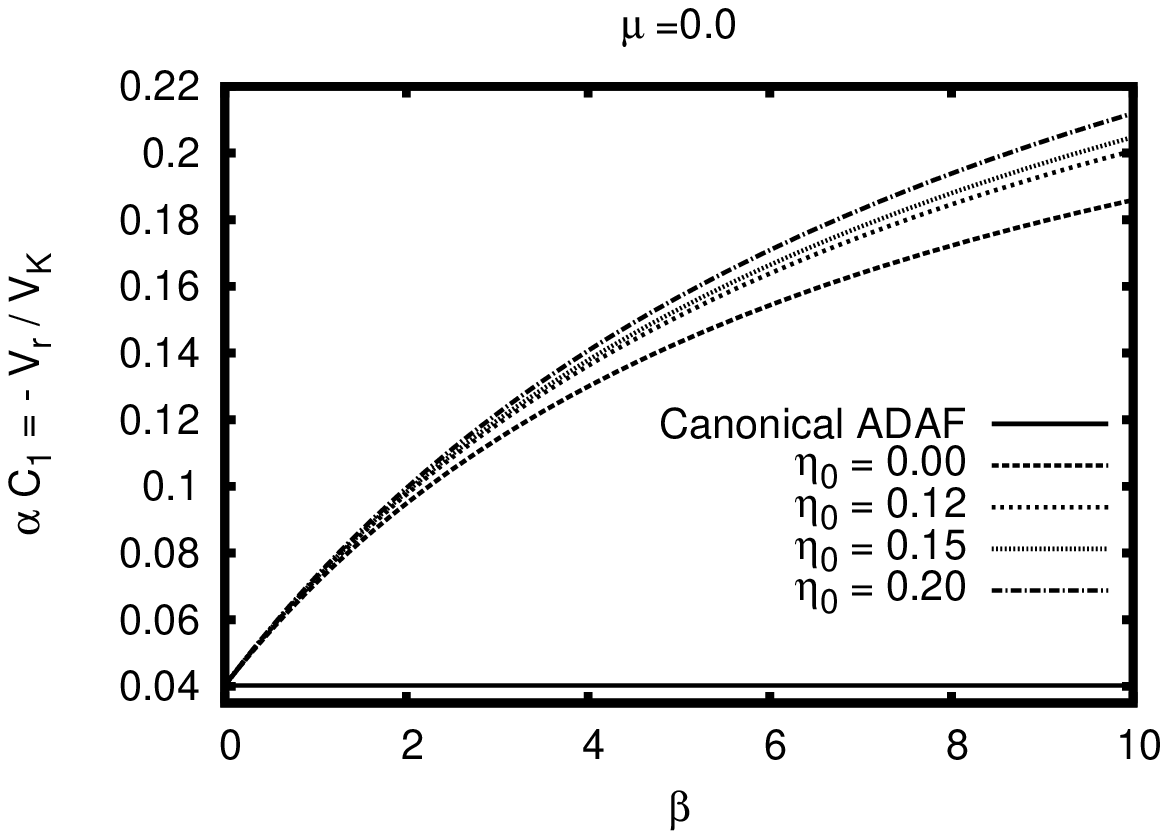}}  {\epsfxsize=7.5cm\epsffile{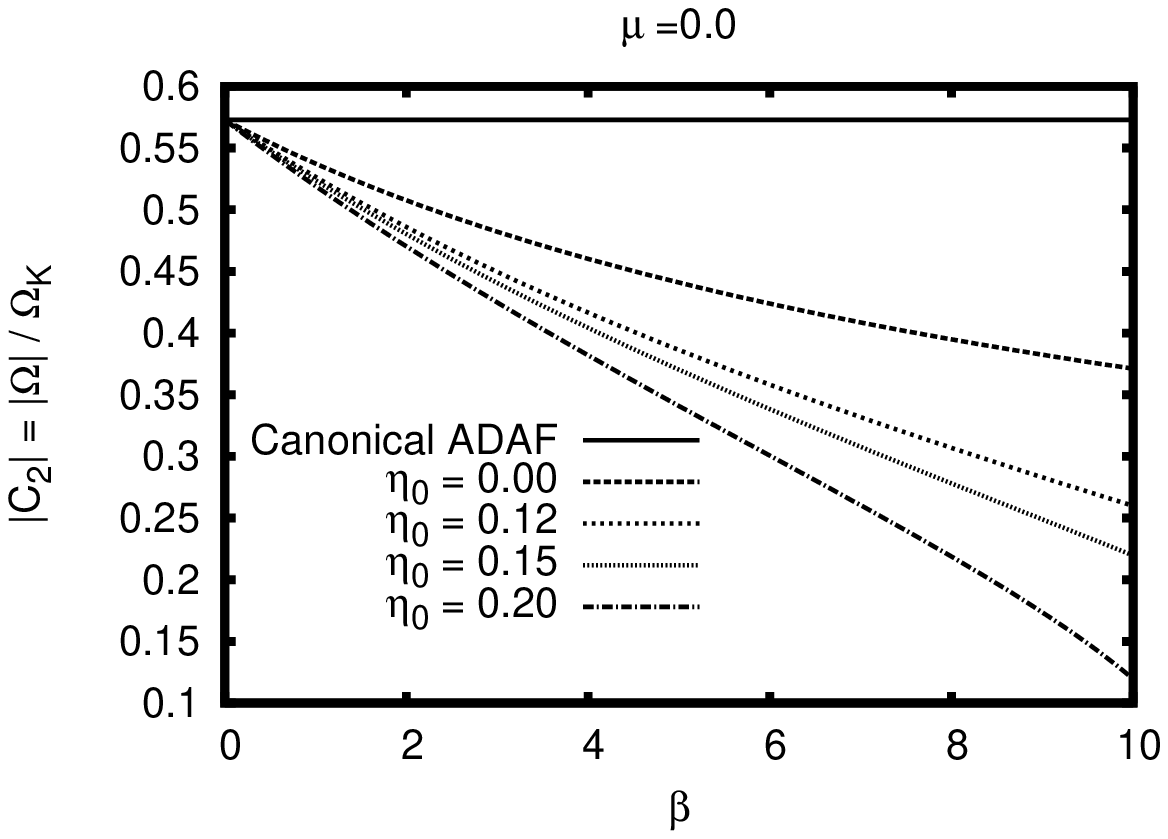}}}
\centerline{{\epsfxsize=7.5cm\epsffile{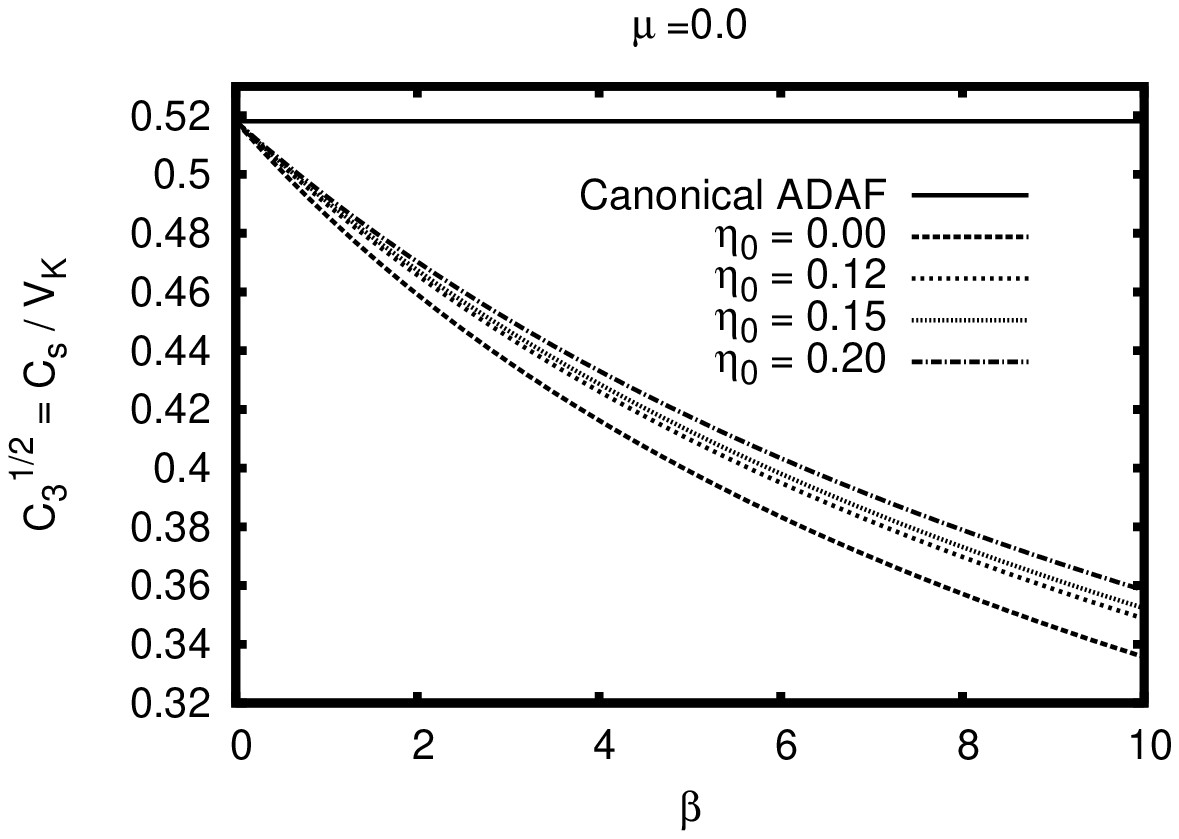}}  {\epsfxsize=7.5cm\epsffile{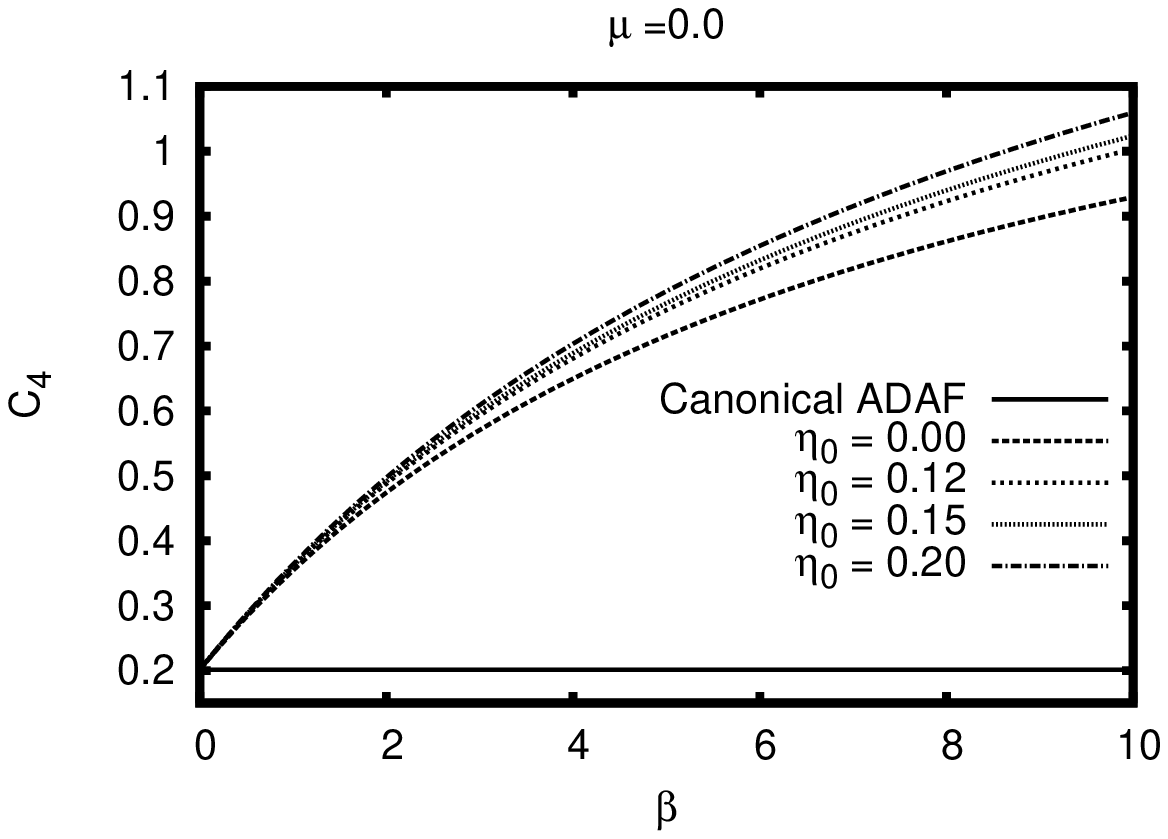}}}
\centerline{{\epsfxsize=7.5cm\epsffile{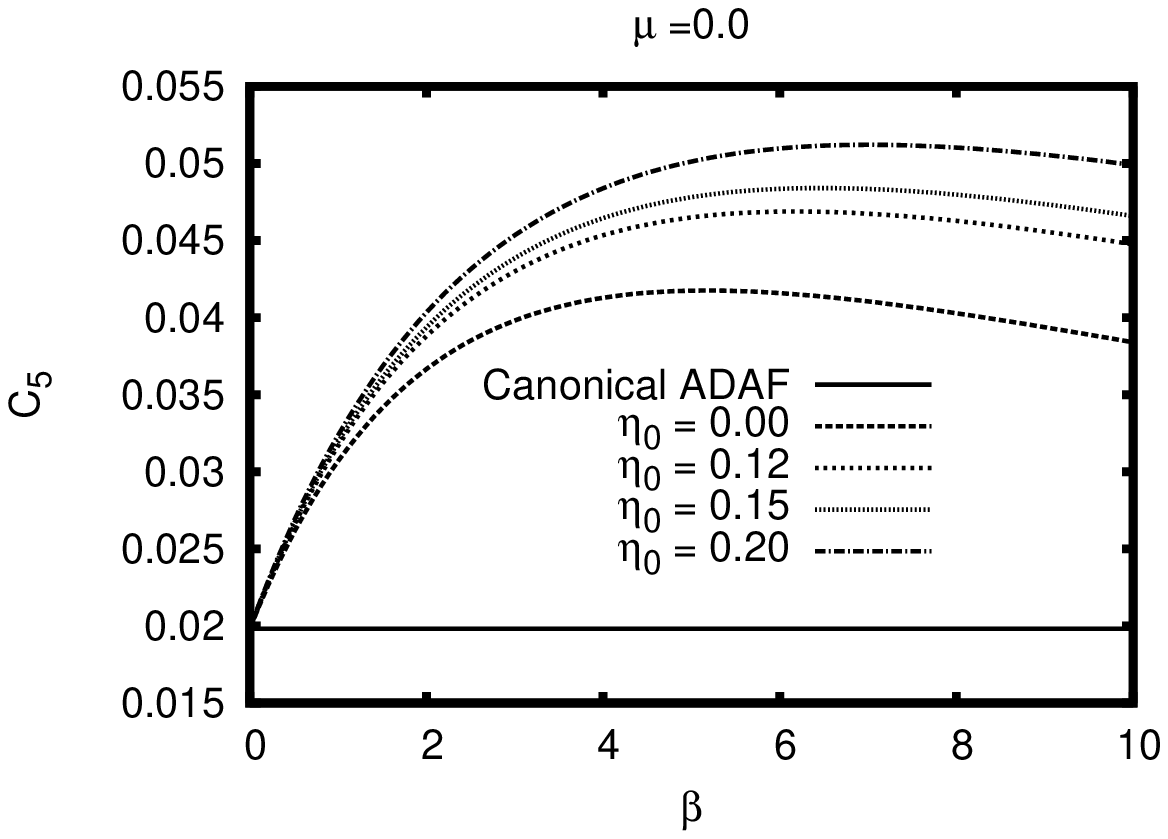}}  {\epsfxsize=7.5cm\epsffile{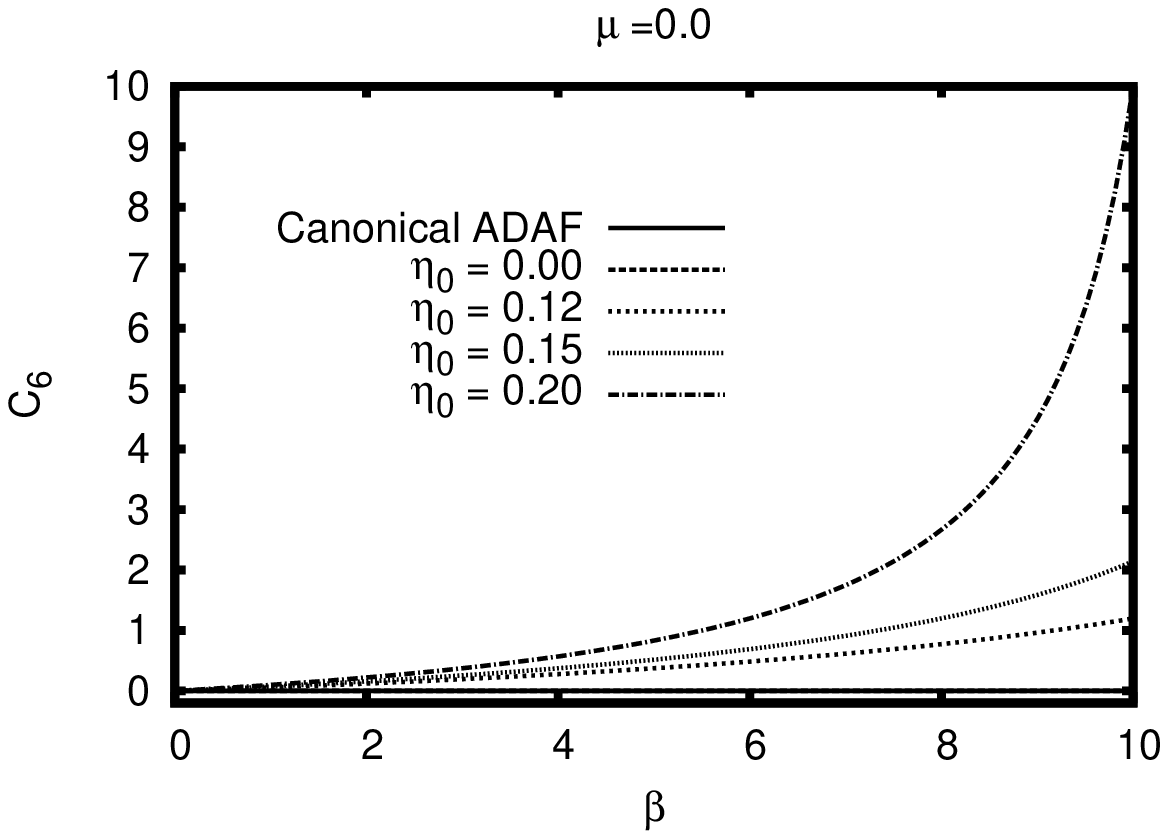}}} 
\end{center}
\vspace{-1.5cm}
\begin{center}
\caption{Physical quantities of the flow as a function of the degree of magnetic pressure to the gas pressure, for several values of $\eta_0=0$, $0.12$, $0.15$, and $0.2$ that corresponding to $P_m=\infty$, $5/6$, $2/3$, and $5/10$. The disc density profile is set to be $s=-3/2$ (no wind), the ratio of the specific heats is set to be $\gamma=1.3$, the viscous parameter is $\alpha=0.1$, and the advection parameter is $f=1.0$.  }
\end{center}
\end{figure}

In this case the function forms of the kinematic coefficient of viscosity and the magnetic diffusivity deviate by factor $(1+\beta)$ from the function forms of used by NY95 and Sh04. Also the profiles of non-resistive and non-magnetic case is shown to compare the present model with canonical ADAF solutions (e.g. NY95). The existence of $(1+\beta)$ in $\nu$ and $\eta$ causes the viscosity and resistivity increase by adding the toroidal magnetic field. The $\nu$ and $\eta$ have direct effects on the viscous torque ($c_4$) and the energy dissipation ($c_5$). Thus, we expect increase of $c_4$ and $c_5$  by adding the toroidal magnetic field that the profiles of $c_4$ and $c_5$ confirm it. Also, by adding the magnetic diffusivity ($\eta_0$) of the flow, $c_4$ and $c_5$ increase. Because the ohmic dissipation increases due to resistivity that it also increases the flow temperature (the profiles of $c_3$ confirm it), and since the  viscous torque is proportional with temperature (sound speed), thus the viscous torque, $c_4$, increases by adding the magnetic diffusivity. The profiles of $c_4$ and $c_5$ imply that for all value of the $\beta$ and $\eta_0$, the total energy dissipation and the viscous torque are larger than the canonical ADAF solutions.  The profiles of $c_3$ show that the temperature of the flow by adding the toroidal component of magnetic field decreases. This property is qualitatively consistent with the results of Bu et al. (2009) and KF09. The profiles of $c_6$ show that the ratio of the resistive dissipation to the viscous dissipation increases by adding the toroidal component of magnetic field ($\beta$) and the magnetic diffusivity ($\eta_0$). As in small amounts of magnetic field and the magnetic diffusivity, the dominant heat generated is the viscous dissipation, while in large values of magnetic field and the magnetic diffusivity, the dominant heat generated will be the resistive dissipation. Also, figure (1) shows by adding the $\beta$ and $\eta_0$ parameters, the radial infall velocity, $c_1$, increases, and the angular velocity, $c_2$, decreases. It can be due to increase of the viscous torque ($c_4$) by parameters of  $\beta$ and $\eta_0$. The raise of the viscous torque by adding $\eta_0$ and $\beta$ parameters generates a larger negative torque in angular momentum equation and causes the angular velocity of the flow decreases, and the matter accretes with larger speed. In the present model, the flow rotates slower than canonical ADAF and accretes speeder than it. The increase of the radial infall velocity by adding the parameter of $\beta$ is qualitatively consistent with the results by Bu et al. (2009) and KF09.

\subsubsection{Second Case: $\mu=1$ }
In the case of $\mu=1$, 
for the kinematic coefficient of viscosity and the magnetic diffusvity we can write
\begin{eqnarray}
 \nonumber \nu=\alpha \frac{c^2_s}{\Omega_K}
\\
=\alpha c_3  \sqrt{G M_*}~r^{1/2}.
\end{eqnarray}
\begin{eqnarray}
 \nonumber \eta=\eta_0 \frac{c^2_s}{\Omega_K}
\\
=\eta_0 c_3 \sqrt{G M_*}~r^{1/2}.
\end{eqnarray}
In this case the function forms of the kinematic coefficient of viscosity and the magnetic diffusivity are the same as Sh04 and NY95 used. The behavior of the physical quantities of the flow in this case are shown in figure (2). The absence of $(1+\beta)$ in this case for $\nu$ and $\eta$ in comparison to previous case causes the value of them decrease by factor $(1+\beta)$. The effects of absence of this factor increases by adding parameter of $\beta$. The profiles of the angular momentum transport ($c_4$) and total energy dissipation ($c_5$) imply that the amounts of them by factor $(1+\beta)$ compared with previous case decrease. As their behavior in terms of the toroidal component of magnetic field have changed and they decrease by adding the $\beta$ parameter. 
However, the magnetic diffusivity has the previous effects and increase these two quantities. Also, the solutions show that the viscous torque and the total dissipation in this case are smaller than canonical ADAF. The weakening of $c_4$ and $c_5$ by adding the $\beta$ parameter reduces the radial infall velocity. Here, the radial infall velocity increases by adding $\eta_0$ that is due to increase of $c_4$ and $c_5$. The behavior of other physical quantities in terms of the $\beta$ and $\eta_0$  parameters does  not change, and represent small variations. To compare the radial infall velocity profiles with canonical ADAF solutions implies that the flow accretes slower than canonical ADAF that is different with previous case. The decrease of the temperature and the viscous torque by adding the parameter of $\beta$ is qualitatively consistent with the results of Bu et al. (2009).

\input{epsf}
\begin{figure}[!ht]
\begin{center}
 \centerline{{\epsfxsize=7.5cm\epsffile{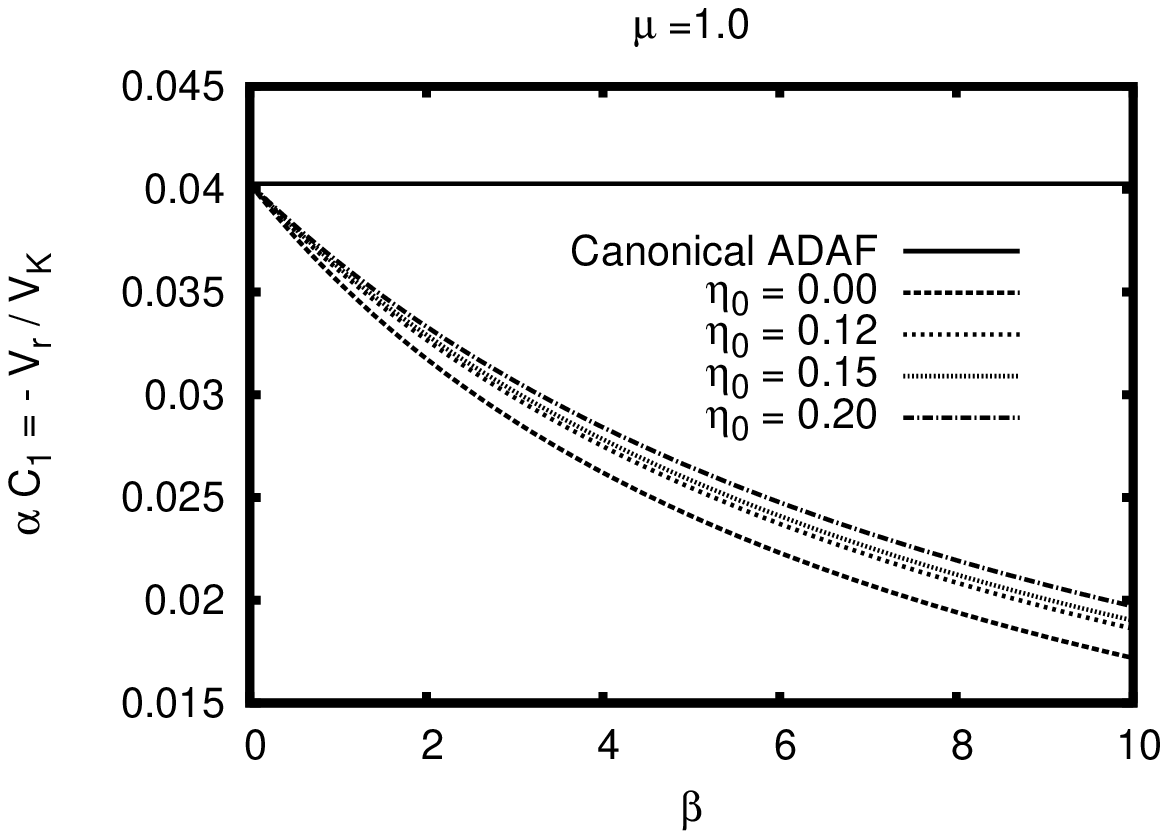}}  {\epsfxsize=7.5cm\epsffile{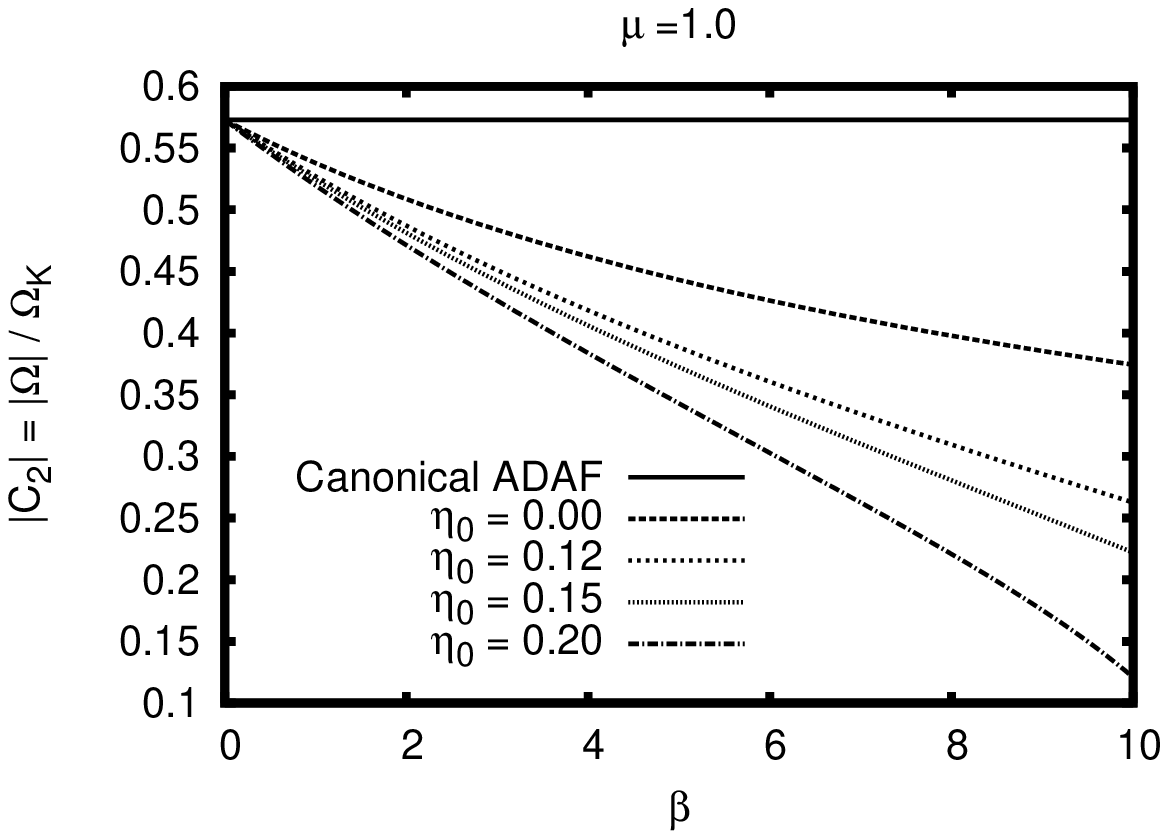}}}
 \centerline{{\epsfxsize=7.5cm\epsffile{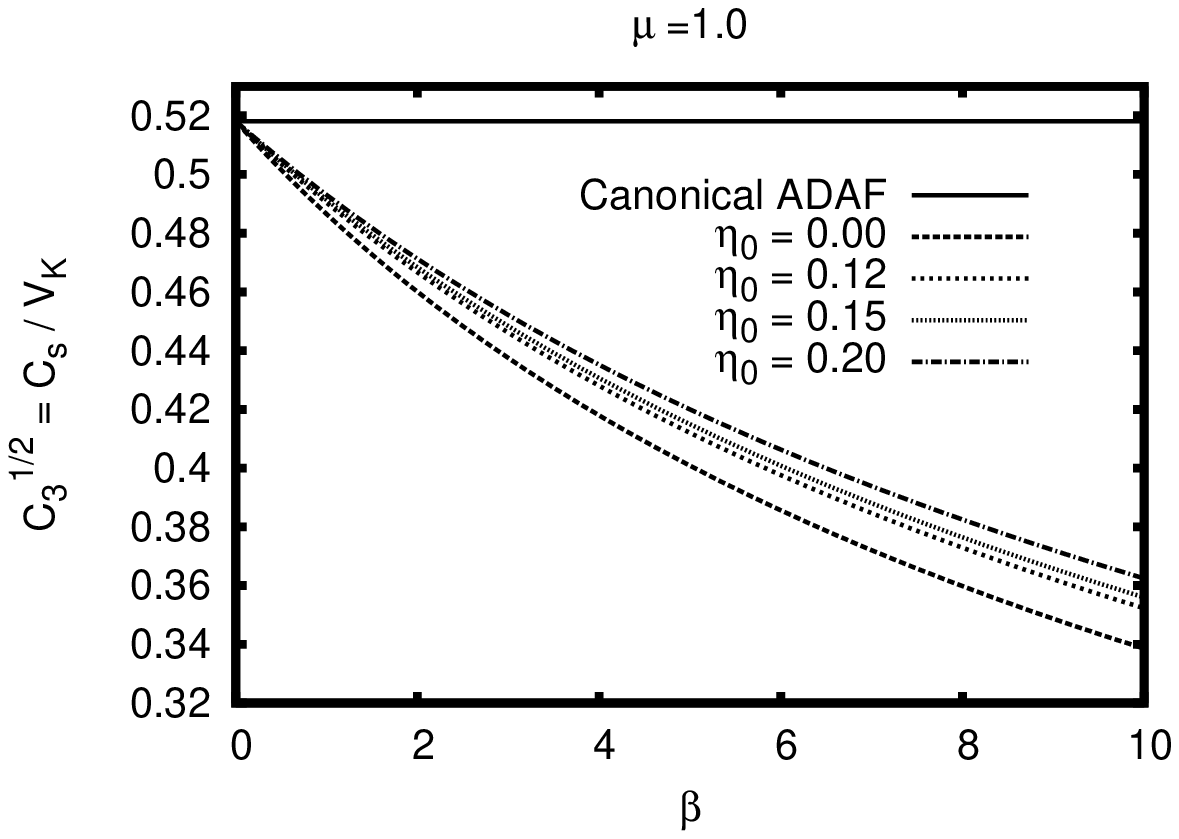}}  {\epsfxsize=7.5cm\epsffile{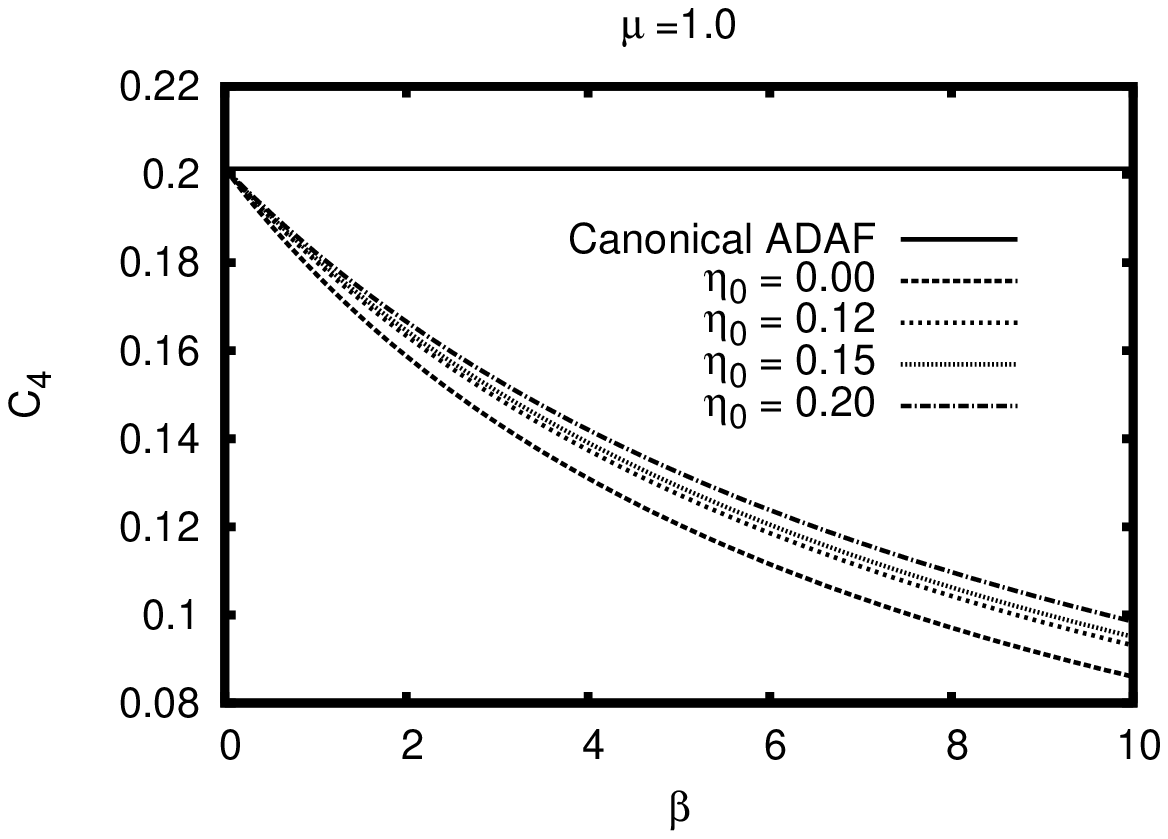}}}
 \centerline{{\epsfxsize=7.5cm\epsffile{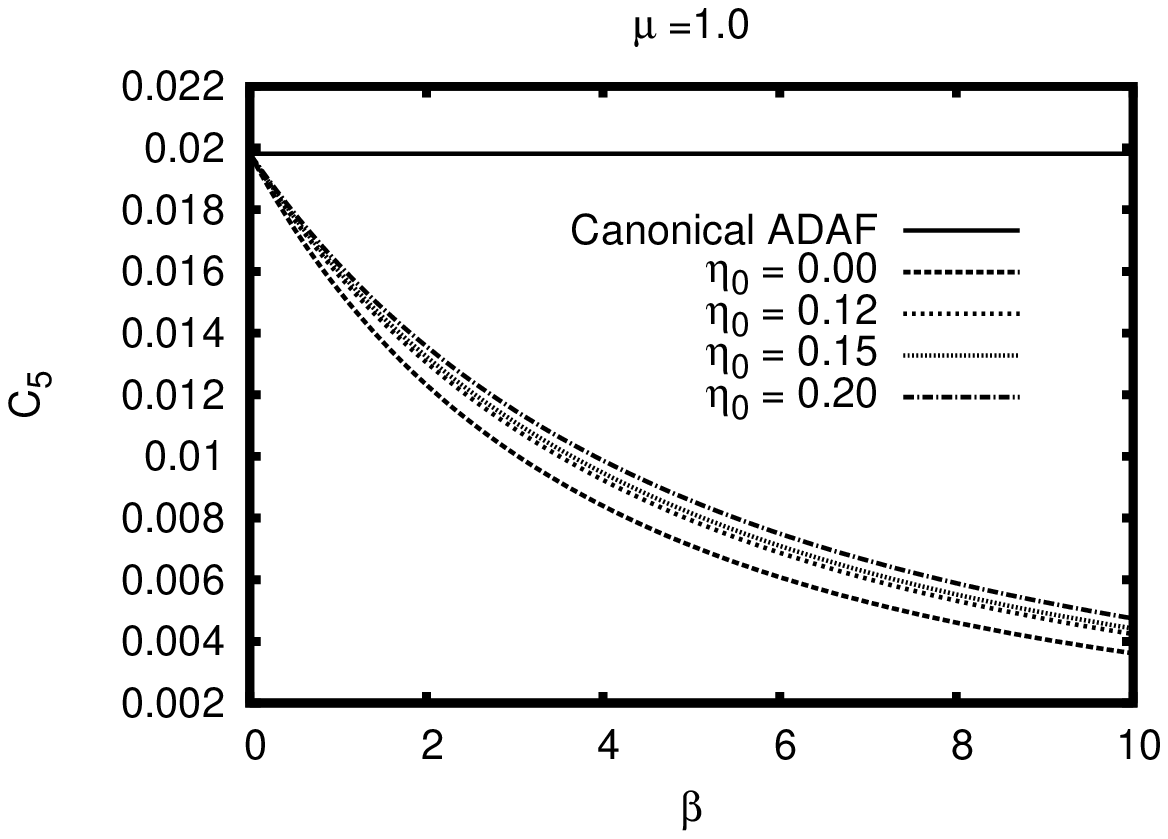}}  {\epsfxsize=7.5cm\epsffile{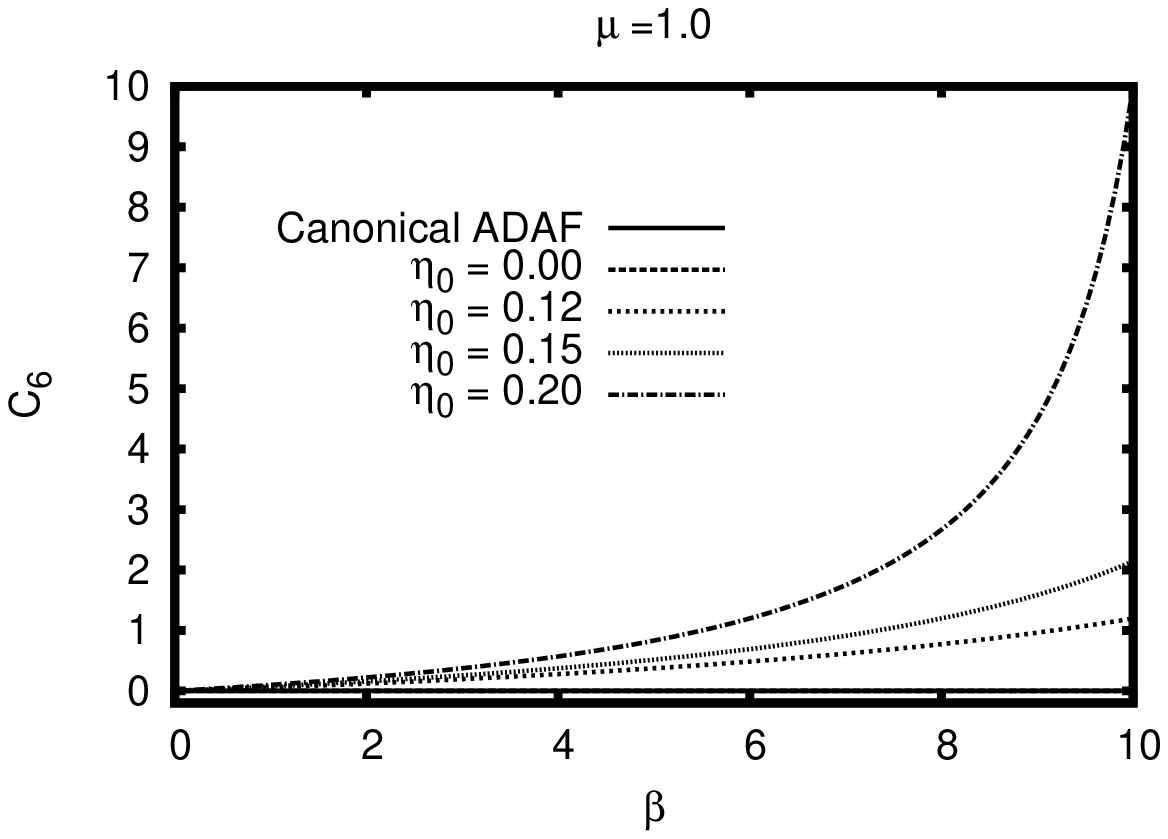}}}
\end{center}
\vspace{-1.5cm}
\begin{center}
\caption{Physical quantities of the flow as a function of the degree of magnetic pressure to 
the gas pressure, for several values of $\eta_0=0$, $0.12$, $0.15$, and $0.2$ that 
corresponding to $P_m=\infty$, $5/6$, $2/3$, and $5/10$. The disc density profile is set 
to be $s=-3/2$ (no wind), the ratio of the specific heats is set to be $\gamma=1.3$, the 
viscous parameter is $\alpha=0.1$, and the advection parameter is $f=1.0$.}
\end{center}
\end{figure}

\subsection{Mass Accretion Rate}
In according to assumptions of section (2) the mass accretion rate defines as
\begin{equation}
\dot{M}=-4\pi r^{2}\rho v_{r}.
\end{equation}
The mass accretion rate under self-similar transformations of equations (21) and (25) becomes
 \begin{equation}
\dot{M}=4\pi\alpha\rho_0 c_1 \sqrt{G M_*}~r^{s+3/2}.
\end{equation}
In our interesting case, $s=-3/2$ (no wind), for the mass accretion rate we can write
 \begin{equation}
\dot{M}=4\pi\alpha\rho_0 c_1 \sqrt{G M_*}
\end{equation}
that implies the mass accretion rate does not vary by position. 
This result is qualitatively consistent with the results by Sh04, Ghanbari et al (2007), and AF06. 
Although the present model of accretion flow is different from Bondi (1952) accretion in various aspect, 
we can define Bondi accretion rate as 
\begin{equation}
 \dot{M}_{Bondi} = \pi G^2 M_*^2 \left(\frac{\rho(\infty)}{c_s^3(\infty)}\right) 
\left[\frac{2}{5-3\gamma}\right]^{(5-3\gamma)/2(\gamma-1)}
\end{equation}
where $\rho(\infty)$ and $c_s(\infty)$ are the density and the sound speed in the gas far away
from the star (Frank et al. 2002). Bondi accretion rate in terms of our self-similar transformations becomes 
\begin{equation}
 \dot{M}_{Bondi} = \pi \sqrt{G M_*} \left(\frac{\rho_0}{c_3^{3/2}}\right)
\left[\frac{2}{5-3\gamma}\right]^{(5-3\gamma)/2(\gamma-1)}.
\end{equation}
Thus, we can write the mass accretion rate in term of Bondi accretion rate as follows
 \begin{equation}
\dot{M}/\dot{M}_{Bondi}=4\alpha c_1 c_3^{3/2}\left[\frac{2}{5-3\gamma}\right]^{(3\gamma-5)/2(\gamma-1)} r^{s+3/2}.
\end{equation}  
In our interesting case, $s=-3/2$ (no wind), we can write
 \begin{equation}
c_7=\dot{M}/\dot{M}_{Bondi}=4\alpha c_1 c_3^{3/2}\left[\frac{2}{5-3\gamma}\right]^{(3\gamma-5)/2(\gamma-1)}.
\end{equation} 
Here, we defined new parameter of $c_7$ that indicates the ratio of the mass accretion rate to Bondi accretion rate. Examples of the coefficient  of $c_7$ in two cases of $\mu=0$ and $1.0$ are shown in figures (3) as a function of the degree of magnetic pressure to the gas pressure, ($\beta$), for different 
value of the magnetic diffusivity, $\eta_0$.  The profiles of $c_7$ show that in the case $\mu=0$, the mass accretion rate increases by adding the toroidal magnetic field and the magnetic diffusivity. While the solutions for $\mu=1$ imply that the mass accretion rate decreases by adding the toroidal magnetic field
and increases by adding the magnetic diffusivity. On the other hand, the magnetic diffusivity in two cases causes the mass accretion rate increase. 
In high magnetic pressure, the $c_7$ profiles for both cases  show that the ratio of the mass accretion rate to the Bondi accretion rate is decreased with
an increase in magnetic pressure. This property is qualitatively consistent with results of Kaburaki (2007). Comparision of the present model with canonical ADAF
solutions implies that in the case of $\mu=0$ the flow accretes quicker than canonical ADAF, however in the case of $\mu=1$ the flow accretes slower than 
canonical ADAF. Also, the profiles of $c_7$  in both of cases show the mass accretion rate in our model is smaller than Bondi accretion rate that is in adapting with observational evidences from Sgr A$^{*}$, M87, and NGC 4261 (Kaburaki 2007).

\input{epsf}
\begin{figure}[!ht]
\begin{center}
 \centerline{{\epsfxsize=7.5cm\epsffile{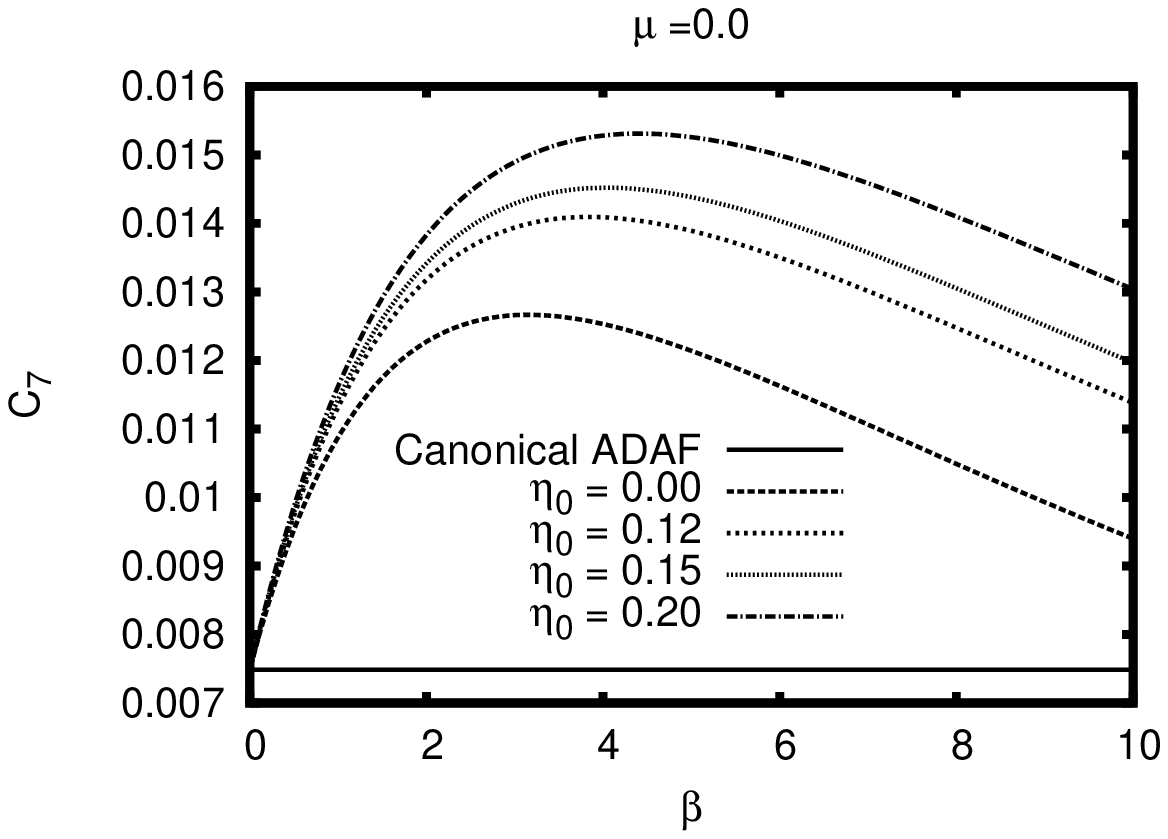}}{\epsfxsize=7.5cm\epsffile{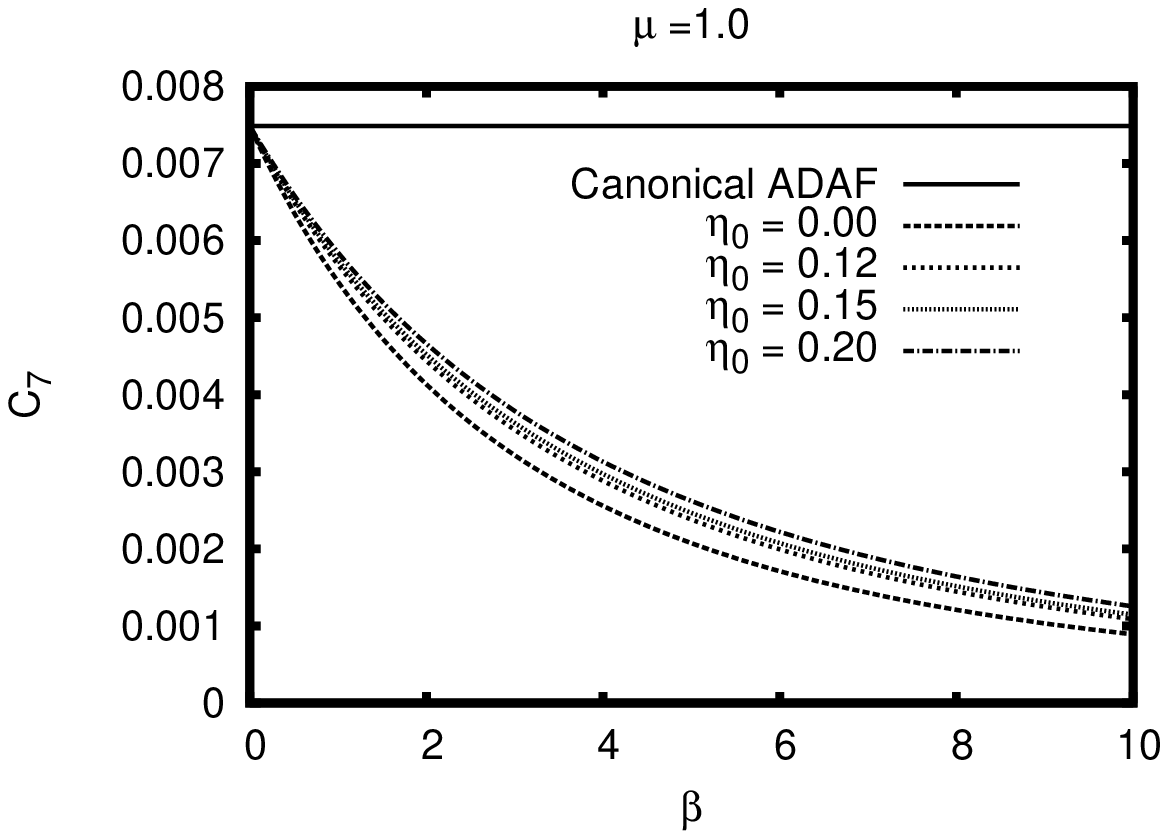}}}  
\end{center}
\vspace{-1.5cm}
\begin{center}
\caption{The ratio of mass accretion rate to Bondi accretion rate ($c_7=\dot{M}/\dot{M}_{Bondi}$) 
as a function of the degree of magnetic pressure to 
the gas pressure, for several values of $\eta_0=0$, $0.12$, $0.15$, and $0.2$ that 
corresponding to $P_m=\infty$, $5/6$, $2/3$, and $5/10$. The disc density profile is set 
to be $s=-3/2$ (no wind), the ratio of the specific heats is set to be $\gamma=1.3$, the 
viscous parameter is $\alpha=0.1$, and the advection parameter is $f=1.0$.}
\end{center}
\end{figure}

In section 3.1, the upper limit of the magnetic field obtained and mentioned with $\beta_b$. 
By substituting $\beta_b$ and equations (33)-(39) in equation (53) and assume of $s=-3/2$, 
the mass accretion rate to the Bondi accretion rate ($c_7$) approximately is 
 \begin{eqnarray}
 \nonumber c_7=\dot{M}/\dot{M}_{Bondi}\approx 24\sqrt{2}~\alpha~ g_1~ \frac{\left(1+\beta_b\right)^{1-\mu}}{\left(5+\beta_b\right)^{5/2}}
\\
= 24\sqrt{2}~\alpha~ g_1~ \frac{\left(1+\frac{18 ~g_2 P_m}{f}\right)^{1-\mu}}{\left(5+\frac{18 ~g_2 P_m}{f}\right)^{5/2}}
\end{eqnarray}
where
\begin{eqnarray}
 \nonumber g_1=\left[\frac{2}{5-3\gamma}\right]^{\frac{(3\gamma-5)}{2(\gamma-1)}}
\end{eqnarray}
\begin{eqnarray}
 \nonumber g_2=\left[\frac{5/3-\gamma}{\gamma-1}\right].
\end{eqnarray}
To obtain the root of derivative of $c_7$ in terms of $f$,  we can write 
\begin{eqnarray}
\nonumber \frac{d c_7}{d f}=0~~\Rightarrow~~ 
\nonumber f_{max}=\left(\frac{18}{5} \right)   \left( \frac{3+2\mu}{1-2\mu} \right) g_2 P_m,
\end{eqnarray} 
The above equation states the mass accretion rate in $f_{max}$ becomes maximum and for $f > f_{max}$ the mass accretion rate decreases by advection degree. 
The $f_{max}$ for $\mu > 1/2$ becomes negative, while $0 \leq f\leq 1$. Thus, the relation of $f_{max}$ is valid only for the case of $\mu=0$. In the case of $\mu=0$, $f_{max}=\left(54/5\right) g_2 P_m$ and $\beta_b=5/3$. Thus, we expect in dominant magnetic case ($\beta > 1$), the accretion efficiency decreases by advection degree parameter. In the case of $\mu=1$, due to the lack of any extremum, the mass accretion rate only increases by advection degree and does not 
show any decrease.

In the case of high magnetic pressure ($\beta \gg 1$), equation (54) becomes
\begin{eqnarray}
 \nonumber c_7=\dot{M}/\dot{M}_{Bondi}\approx 24\sqrt{2}~\alpha~ g_1~ \left(\frac{f}{18 ~g_2 P_m}\right)^{3/2+\mu}.
\end{eqnarray} 
The above equation implies that the mass accretion rate to Bondi accretion is strongly depends on viscosity parameter, 
Prandtl number and the advection degree. The mass accretion rate increases by $f$ and decreases by $P_m$. 
Thus, accretion efficiency increases by the advection degree parameter in high magnetic pressure.


\subsection{Timescales}
To estimate the effect of viscosity and resistivity on the accretion
discs, we compare the viscous and resistive timescales with accretion timescale. The accretion timescale, $t_{acc}$,
and the viscous timescale, $t_{visc}$, are given by
  \begin{equation}
\nonumber t_{acc}=\frac{r}{-v_r},
\end{equation} 
  \begin{equation}
\nonumber t_{visc}=\frac{r^2}{\nu}.
\end{equation}
 We are using a similar functional form of $t_{visc}$ for the resistive timescale, $t_{resis}$, that is given by
  \begin{equation}
\nonumber t_{resis}=\frac{r^2}{\eta}.
\end{equation}
By using self-similar forms of physical quantities, we can write
  \begin{eqnarray}
\nonumber \frac{t_{resis}}{t_{acc}}=\frac{\alpha}{\eta_0}\frac{c_1}{c_3}(1+\beta)^{\mu-1}
\\ 
=3(\frac{\alpha}{\eta_0})(s+2).
\end{eqnarray}
The equation (30) is used for fraction of $c_1/c_3$. As, we said in previous section, in present model $\alpha/\eta_0$ 
is the magnetic Prandtl number, $P_m$, so above equation becomes
 \begin{equation}
\nonumber \frac{t_{resis}}{t_{acc}}=3 P_m (s+2).
\end{equation}
For our interesting case, $s=-3/2$ (no wind), we can write
 \begin{equation}
\nonumber \frac{t_{resis}}{t_{acc}}=\frac{3}{2} P_m .
\end{equation} 
The above equation implies that for $P_m \le 2/3$, the magnetic diffusivity timescale is shorter than or equal to 
accretion timescale, while for $P_m > 2/3$ the accretion timescale is shorter. 
\\ Similar calculations for the viscous timescale express
\begin{equation}
\nonumber \frac{t_{visc}}{t_{acc}}=3 (s+2),
\end{equation}
where in no wind case ($s=-3/2$) becomes
\begin{equation}
\nonumber \frac{t_{visc}}{t_{acc}}=(3/2).
\end{equation}
Thus, the viscosity timescale will be longer than the accretion timescale. To compare 
the magnetic diffusivity with the viscous timescales, we can write
\begin{eqnarray}
\nonumber \frac{t_{resis}}{t_{visc}}=\frac{r^2/\eta}{r^2/\nu}
\\
\nonumber=\frac{\nu}{\eta}~~
\\
=P_m.
\end{eqnarray}    
Thus, the magnetic Prandtl number specifies which one is shorter. For example in flow with high conductivity (e.g. AF06; KF09), 
$\eta\rightarrow 0$, the magnetic Prandtl number limits to infinity, and so the magnetic diffusivity timescale will be very longer
than the viscous timescale. On the other hand, for a flow with finite resistivity and tiny viscosity (e.g. Sh04), 
the magnetic Prandtl number limits to zero, and so the magnetic diffusivity timescale is very shorter than viscous timescale. When
the resistivity and the viscosity are approximately equal, $P_m\sim1$, we expect $t_{resis}\sim t_{visc}$. Also, in 
special case of $P_m=5/6$ and $s=-3/2$ that escape and creation of magnetic field 
are balanced and there is no mass-loss, $t_{resis}= (5/6) t_{visc}$.

\section{Summary and Discussion}
In this paper, the influences of the resistivity on the structure of the advection-dominated accretion flow is investigated.
It is used only azimuthal component of magnetic field that is consistent with observational
evidence of Galactic center (Novak et al 2003; Chuss et al. 2003; Yuan 2006).
The $\alpha$-prescription is used for the kinematic coefficient of viscosity and the magnetic diffusivity. 
The equations of the model are solved by a semi-analytical self-similar method in comparison with the 
self-similar solution by AF06. 

The physical quantities of disc are sensitive to the amounts of 
the magnetic pressure fraction ($\beta$) and the magnetic diffusivity ($\eta_0$) parameters. 
As, the angular velocity of the flow by adding the $\beta$ and $\eta_0$ parameters decreases.
For a value of the magnetic pressure fraction, the angular 
velocity of disc becomes zero. This amount of the magnetic pressure fraction strongly depends on 
the properties of the accreting gas, such as the viscosity, resistivity, adiabatic index, and advection degree.
The solutions represent the radial infall velocity increases by adding the magnetic diffusivity.
Also the solutions show that the temperature of the flow decrease by adding the toroidal component of magnetic field.
This result qualitatively is consistent with the results of Bu et al. (2009) and KF09.
The profiles of the temperature of the flow show that it increases by adding the
magnetic diffusivity that is due to the raise of the resistive dissipation. 
Comparison of the present model with Bondi accretion
implies that for all values of the $\beta$ and $\eta_0$ parameters, the Bondi accretion rate is
larger than the mass accretion rate  
 that is in accord with observational evidences of Sgr A$^{*}$, M87, and NGC 4261 (Kaburaki 2007). 
Also, the mass accretion rate profiles at 
high magnetic field express that the magnetic field reduces the mass accretion rate that is similar to 
results of Kaburaki (2007).
We found that in the small magnetic field, the more heat generated in 
the flow is due to the viscous dissipation, while the ohmic dissipation will be dominant in large amounts of magnetic field and resistivity. 

As noted in the introduction, the MHD simulations show that linear growth of MRI decreases significantly by ohmic dissipation. 
linear growth of the MRI in the resistive fluid can be characterized by the Lundquist number  ($S_{MRI}=c_A^2/\eta\Omega$) and 
magnetic Reynolds number ($Re_{M}=c_s^2/\eta\Omega$), where $c_A$, $c_s$, $\eta$, and $\Omega$ have usual meaning. 
In terms of our self-similar transformations, the Lundquist number and magnetic Reynolds number become 
$S_{MRI}=2\beta/\eta_0 c_2 (1+\beta)^{1-\mu}$ and $Re_{M}=1/\eta_0 c_2 (1+\beta)^{1-\mu}$. The solutions of
 present model show that $S_{MRI}$ and  $Re_{M}$ decrease by resistivity. This property is qualitatively 
consistent with MHD simulation results (Fleming et al. 2000; Masada \& Sano 2008).

Here, latitudinal dependence of physical quantities is ignored,
while some authors showed that latitudinal dependence is
important in structure consideration of a disc (Narayan \& Yi 1995; Sh04; Ghanbari et al. 2007). 
One can investigate latitudinal behavior of such discs. 
Furthermore, in a realistic model the 
advection parameter $f$ is a function of position, one can consider such discs.

\section*{Acknowledgements}
I would like to thank the referee for very useful comments that helped me to improve the initial version of the paper.
I would also like to thank Markus Flaig for helpful discussion.

\end{document}